\documentclass[12pt,a4paper,twoside]{article}
\usepackage{epsfig}
\usepackage{baltlat1}
\usepackage{wrapfig}
\def\k13{\mbox{$\kappa_{\mbox{\tiny 1.3mm}}$}}

\def\kms{\rm{km\, s^{-1}}}

\def\as{\mbox{$a_{\mbox{\tiny s}}$}}
\def\nhd{\mbox{$n_{\mbox{\tiny H}_{2}}$}}

\def\>{$>$}
\def\<{$<$}

\def\ltsima{$\; \buildrel < \over \sim \;$}
\def\simlt{\lower.5ex\hbox{\ltsima}}
\def\gtsima{$\; \buildrel > \over \sim \;$}
\def\simgt{\lower.5ex\hbox{\gtsima}}
\def\arcsec{\hbox{$^{\prime\prime}$}}

\def\h2{\mbox{$_{\mbox{\tiny H2}}$}}

\begin{document}
\ \
\vspace{0.5mm}

\setcounter{page}{1}
\vspace{8mm}

\titlehead{Baltic Astronomy, vol.~12, XXX--XXX, 2003.}

\titleb{Detailed Studies of Cloud Cores: Probing the Initial Conditions for Protostellar Collapse}

\begin{authorl}
\authorb{Ph.~Andr\'e}{1}, 
\authorb{A.~Belloche}{2},
\authorb{P.~Hennebelle}{3} and
\authorb{D.~Ward-Thompson}{4}
\end{authorl}

\begin{addressl}
\addressb{1}{CEA Saclay, Service d'Astrophysique, 
  Orme des Merisiers, B\^at.~709, F-91191 Gif-sur-Yvette Cedex, France}

\addressb{2} {Max-Planck-Institut f\"ur Radioastronomie, Auf dem H\"ugel 69,
D-53121 Bonn, Germany}

\addressb{3} {Laboratoire de Radioastronomie, ENS, 24 rue Lhomond, 
F-75231 Paris Cedex 05, France}

\addressb{4} {Dept of Physics, Cardiff University, PO Box 913, Cardiff, UK}

\end{addressl}

\submitb{Received December 4, 2003}

\begin{abstract}
Improving our understanding of the earliest stages of star formation 
is crucial to gain insight into the origin of stellar masses, 
multiple systems, and protoplanetary disks.
We discuss recent advances made in this area thanks to detailed 
mapping observations at infrared and (sub)millimeter wavelengths. 
Although ambipolar diffusion appears to be too slow
to play a direct role in the formation of dense cores, 
there is nevertheless good evidence that the gravitational collapse 
of isolated protostellar cores is strongly magnetically controlled. 
We also argue that the beginning of 
protostellar collapse is much more violent in cluster-forming clouds than
in regions of distributed star formation. 
\end{abstract}


\resthead{Detailed Infrared and Submillimeter Studies of Cloud Cores}{Ph. Andr\'e, 
A.~Belloche, P.~Hennebelle, D.~Ward-Thompson}



\sectionb{1}{INTRODUCTION}

One of the main limitations in our present understanding of the star formation 
process
is that we do not know well the initial conditions for protostellar collapse.
In particular, there is a major controversy at the moment 
between two schools
of thought for the formation and evolution of dense cores within 
molecular clouds: The classical picture based on magnetic support and 
ambipolar diffusion (e.g. Shu et al. 1987, 2003; 
Mouschovias \& Ciolek 1999) 
has been seriously challenged by a new, more dynamical picture, which 
emphasizes the role of supersonic turbulence in supporting clouds on large
scales and generating density fluctuations on small scales 
(e.g. Klessen et al. 2000; Padoan \& Nordlund 2002).
Improving our knowledge of the earliest stages of star formation 
is thus of prime importance, especially
since there is now good evidence that these stages control the origin of 
the stellar initial mass function (e.g. Motte, Andr\'e, \& Neri 1998; 
Testi \& Sargent 1998; Johnstone et al. 2000). 

Here, we discuss recent observational advances concerning the density and 
velocity structure of cloud cores (\S ~2 and \S ~3), and we compare the 
results with the predictions of  theoretical models (\S ~4).


\sectionb{2}{DENSITY STRUCTURE OF PRESTELLAR CORES}

Two techniques have been used to trace the density structure 
of cloud cores (see Fig.~1a): (1)~mapping the optically thin 
(sub)millimeter continuum {\it emission} from the cold dust contained 
in the cores, and (2) mapping the same cold core dust in {\it absorption} 
against the background infrared emission (originating from warm cloud dust 
or remote stars). 
 
Ward-Thompson et al. (1994, 1999) and Andr\'e et al. (1996)  
employed the first approach to probe the structure of prestellar cores
(see also Shirley et al. 2000).
Under the simplifying assumption of spatially uniform dust temperature and
emissivity properties, they concluded that the radial density profiles of
isolated prestellar cores were {\it not} consistent with the single 
$\rho(r) \propto r^{-2}$ power law of the singular isothermal sphere
(SIS) but were flatter than 
$\rho(r) \propto r^{-1}$ in their inner regions (for $r \leq R_{flat}$), 
and approached $\rho(r) \propto r^{-2}$ only beyond a typical radius
$R_{flat} \sim $~2500--5000~AU. 

More recently, the use of the {\it absorption} approach, 
both in the mid-IR from space (e.g. Bacmann et al. 2000, 
Siebenmorgen \& Kr\"ugel 2000) and in the near-IR from the ground 
(e.g. Alves et al. 2001), made it possible to confirm and extend the 
(sub)millimeter emission results, essentially independently of any assumption
about the dust temperature distribution. 
 
The typical column density profile found by these emission and 
absorption studies of prestellar cores 
has the following characteristics (see, e.g., Fig.~1b):

\noindent
a) a flat inner region (of radius $R_{flat} = 5000 \pm 1000$~AU for L1689B 
according to the fitting analysis shown in Fig.~1b), 
b) a region roughly consistent with $N\h2 \propto \bar{r}^{-1}$ 
(corresponding 
to $\rho \propto r^{-2}$ for a spheroidal core), 
c) a sharp edge where the column density falls off more rapidly than 
$N\h2 \propto \bar{r}^{-2}$ with projected radius 
defining the core outer radius ($R_{out} = 28000 \pm 1000$~AU for L1689B).

\begin{figure}
\vskip2mm
\centerline{{\psfig{figure=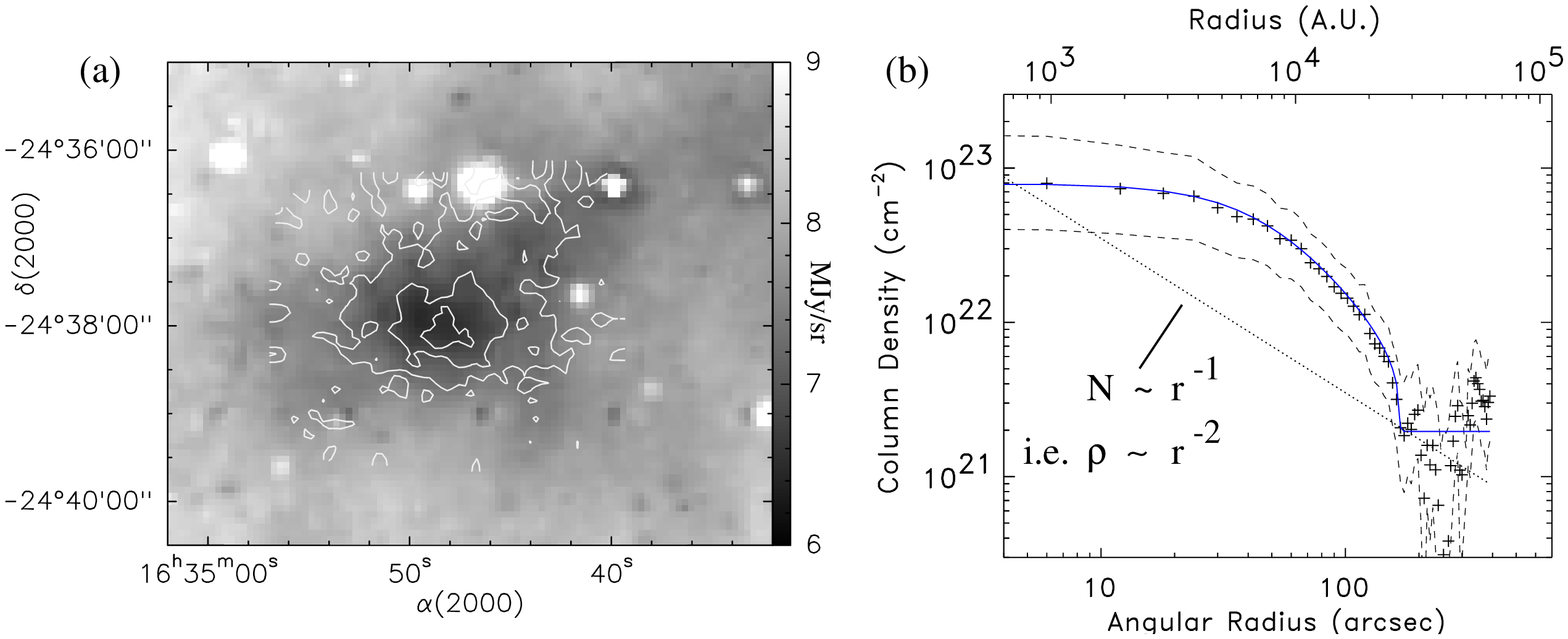,height=13.cm,angle=270,clip=}}}
\captionb{1}{(a) ISOCAM 6.75~$\mu$m absorption image of the prestellar core 
L1689B in Ophiuchus (scale on the right in MJy/sr). 
The 1.3~mm continuum emission map of André et al. (1996) with the IRAM 30~m telescope is superposed as contours 
(levels: 10, 30, 50~mJy/13\arcsec -beam).
(b) Column density profile of L1689B (crosses)
derived from the absorption map shown in (a) by averaging the intensity over 
elliptical annuli for a 40$^\circ $ sector in the southern part of the core. 
The dashed curves show the most extreme profiles compatible with the data 
given the uncertainties  
affecting the absorption analysis.
The solid curve is the best fit of a Bonnor-Ebert sphere
model (embedded in a medium of uniform column density), 
obtained with the following parameters:
$\rho_c/\rho_{out} = 40 \pm 15$ (i.e., well into the 
unstable regime), 
$T_{eff} = 50 \pm 20$~K, 
$P_{ext}/k_B = 5\pm 3 \times 10^5\, \rm{K}\, \rm{cm}^{-3} $.
For comparison, the dotted line shows the $N\h2 \propto \bar{r}^{-1}$ 
profile of a SIS at $T = 10$~K.  
(Adapted from Bacmann et al. 2000.)
}
\vskip2mm
\label{fig-abs}
\end{figure}

\subsectionb{2.1}{Comparison with models of core structure}

The results summarized above 
set strong constraints on the density structure at the onset of 
protostellar collapse.
The circularly-averaged column density profiles 
can often be fitted remarkably well with models of pressure-bounded 
Bonnor-Ebert spheres, as first demonstrated by Alves et al.~(2001) for B68. 
This is also the case of L1689B as illustrated in Fig.~1b. 
The quality of such fits shows that 
Bonnor-Ebert spheroids provide a good, first order model for the structure of
isolated prestellar cores. In detail, however, there are several problems with
these Bonnor-Ebert models. 
First, the inferred density contrasts (from center to edge) 
are generally larger (i.e., $\simgt $~20--80 -- see Fig.~1b) 
than the maximum contrast of $\sim $~14 for stable Bonnor-Ebert 
spheres.
Second, the effective core temperature needed in these models 
(for thermal pressure gradients to balance self-gravity)
is often significantly larger than both the average dust temperature 
measured with ISOPHOT (e.g. Ward-Thompson et al. 2002) 
and the gas temperature measured in NH$_3$ 
(e.g. Lai et al. 2003). In the case of 
L1689B, for instance, the effective temperature of the Bonnor-Ebert fit 
shown in Fig.~1b is $T_{eff} \sim 50$~K, while the dust temperature
observed with ISOPHOT is only $T_d \sim 11$~K (Ward-Thompson et al. 2002).
Third, the physical process responsible for bounding the cores at some
external pressure is unclear. These arguments suggest that prestellar 
cores cannot simply be described as isothermal 
hydrostatic structures and are either already contracting 
(e.g. Lee, Myers, Tafalla 2001) 
or experiencing extra support from static or turbulent magnetic fields 
(e.g. Curry \& McKee 2000). 
It has also recently been pointed out that good Bonnor-Ebert fits can often 
be found for non-equilibrium, transient ``cores'' produced by turbulence (Ballesteros-Paredes et al. 2003).

As shown by Bacmann et al. (2000), 
one way to account for large density contrasts and high effective
temperatures is to consider models of cores initially supported 
by a static magnetic field and evolving through ambipolar diffusion 
(e.g. Ciolek \& Mouschovias 1994, Basu \& Mouschovias 1995). 
A serious problem, however, with such models 
is that they need a strong   
field {\it in the low-density ambient cloud} 
($\sim $~30--100~$ \mu$G)  to explain the observed 
sharp edges (see Bacmann et al. 2000).  
This seems 
inconsistent with 
existing Zeeman measurements 
(e.g. Crutcher 1999, Crutcher \& Troland 2000). 
Furthermore, when the ambient field is strong (highly subcritical case),
the model cores evolve on the ambipolar 
diffusion timescale $t_{AD} \sim 10\, t_{ff}$ (Ciolek \& Basu 2001; 
$t_{ff}$ is the free-fall time), while observed cores have significantly
shorter lifetimes $\sim 4\, t_{ff}$ at $\nhd \sim 10^4\, \rm{cm}^{-3}$ 
(cf. Jessop \& Ward-Thompson 2000). On the other hand, isolated starless cores 
appear to live $\simgt 3$ times {\it longer} than typical density 
fluctuations generated by turbulence 
(cf. Ballesteros-Paredes et al. 2003.) 
It is possible that more elaborate ambipolar diffusion 
models, incorporating the effects of a non-static, turbulent magnetic field 
in the outer parts of the cores and in the ambient cloud, 
would be more satisfactory and could 
also account for the filamentary shapes often seen on large 
($\simgt 0.25$~pc) scales (cf. Jones \& Basu 2002).

\sectionb{3}{VELOCITY STRUCTURE OF CLASS 0 ENVELOPES}


Another approach to constraining the collapse initial conditions consists
in observing the detailed properties of very young accreting protostars 
(Class~0 objects -- e.g. Andr\'e, Ward-Thompson, Barsony 2000) that 
still retain detailed memory of their genesis.
There have been very few quantitative studies of the velocity structure 
of the envelopes of Class~0 objects so far. Here, we describe two recent, 
contrasted examples.

\vskip1mm
\begin{wrapfigure}{i}[0pt]{61mm}
\centerline{\psfig{figure=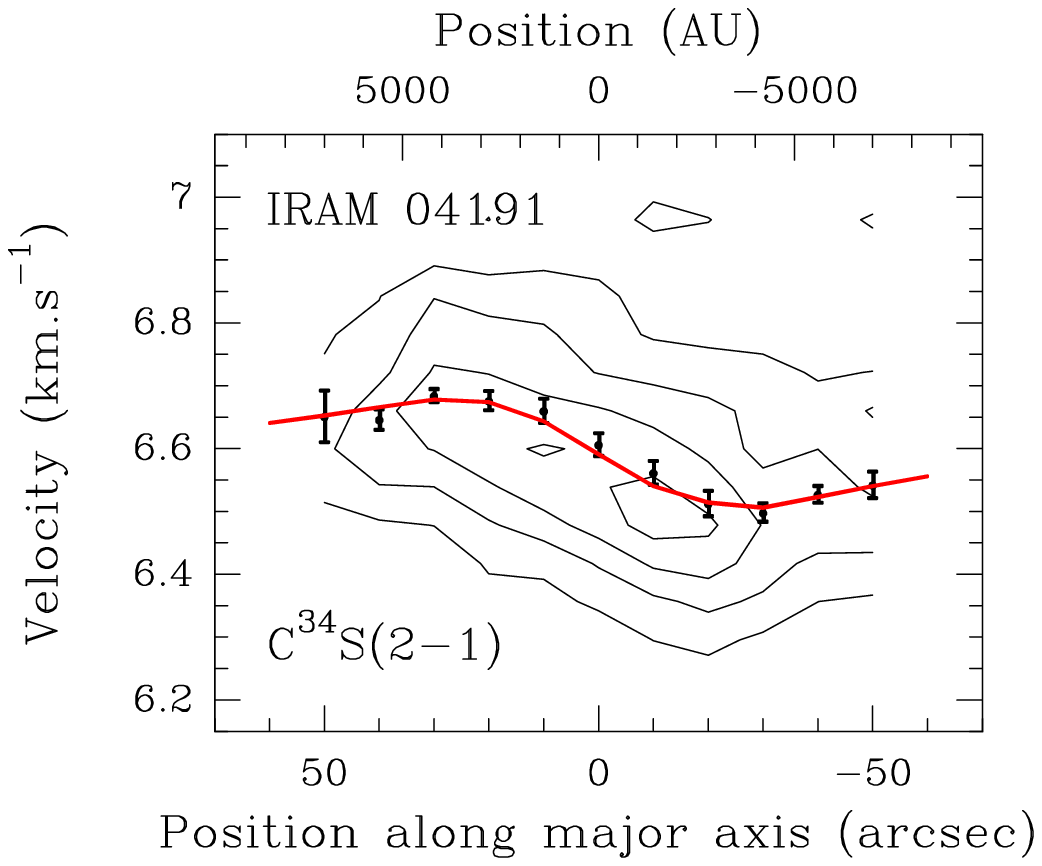,width=50truemm,angle=270,clip=}}
\captionb{2}{Position-velocity diagram along the major axis of the  
IRAM~04191 envelope (i.e., perpendicular to the flow) 
based on a C$^{34}$S$(2-1)$ 
map taken at the IRAM 30m telescope (Belloche et al. 2002).
Contours: 0.2 to 0.8 by 0.2~K.
The dots with error bars mark the observed velocity centroids. 
The solid curve shows the profile of a model with differential 
rotation [$V_{rot}(r >3500\, \rm{AU} ) \propto r^{-1.5}$].}
\label{rotation}
\end{wrapfigure}

The first object, IRAM~04191$+$1522 (IRAM~04191 for short), 
is relatively {\it isolated}. It was found at 1.3~mm by 
Andr\'e, Motte, \& Bacmann (1999) in the southern part of the Taurus cloud
($d = 140$~pc).
IRAM~04191 is associated with a prominent, flattened protostellar 
envelope, seen in the (sub)millimeter dust continuum and in
dense gas tracers such as N$_2$H$^+$, C$_3$H$_2$, H$^{13}$CO$^+$, 
and DCO$^+$ (Belloche et al. 2002). 
All of the maps taken at the IRAM 30m telescope in 
small optical depth lines show a clear 
rotational velocity gradient across the envelope of 
$\sim 9\, \kms \,$pc$^{-1}$ (after deprojection), 
perpendicular to the outflow axis. 
This gradient is one order of magnitude
larger than those typically observed in starless cores 
(e.g. Goodman et al. 1993). Furthermore, the rotation of the protostellar 
envelope does not occur in a rigid-body, but differential, fashion: 
the inner $\sim $~3500~AU-radius region rotates significantly faster than 
the outer parts of the envelope, as indicated by the characteristic ``S'' 
shape of the position-velocity diagrams obtained by 
Belloche et al. (see, e.g., Fig.~2). 
Fast, differential rotation is expected in protostellar envelopes because 
of conservation of angular momentum during dynamical collapse. In the present
case, however, the dramatic drop in rotational velocity observed at 
$r \geq 3500$~AU (Fig.~3a), combined with the flat infall 
velocity profile (see below), points to {\it losses of angular momentum} in the 
outer envelope (see discussion in \S ~4 below). 

Direct evidence for infall motions over a large portion of the 
envelope of IRAM~04191 is observed in optically thick lines such as 
CS(2--1), CS(3--2), 
H$_2$CO($2_{12} - 1_{11}$), and H$_2$CO($3_{12} - 2_{11}$). 
These lines are double-peaked and skewed to the blue up to 
$\simgt $~40\arcsec ~from source center, which is 
indicative of infall motions up to a radius 
$R_{inf} \simgt 5000$~AU (cf. Evans 1999, 2003). 
Radiative transfer modeling confirms this view, suggesting a  
flat, subsonic infall velocity profile ($V_{inf} \sim 0.1\, \kms $) for 
$3000 \simlt r \simlt 11000$~AU and larger infall velocities scaling 
as $V_{inf} \propto r^{-0.5}$ for $r \simlt 3000$~AU
(see Fig.~3b and Belloche et al. 2002 for details).
The mass infall rate is estimated to be 
$\dot{M}_{inf} \sim 2-3 \times a_s^3/G 
\sim 3 \times 10^{-6}$~M$_\odot$~yr$^{-1}$ 
(where $a_s \sim 0.15-0.2\, \kms $ is the sound speed
at $T \sim 6-10$~K), 
roughly independent of radius.

\begin{figure}
\vskip2mm
\centerline{{\psfig{figure=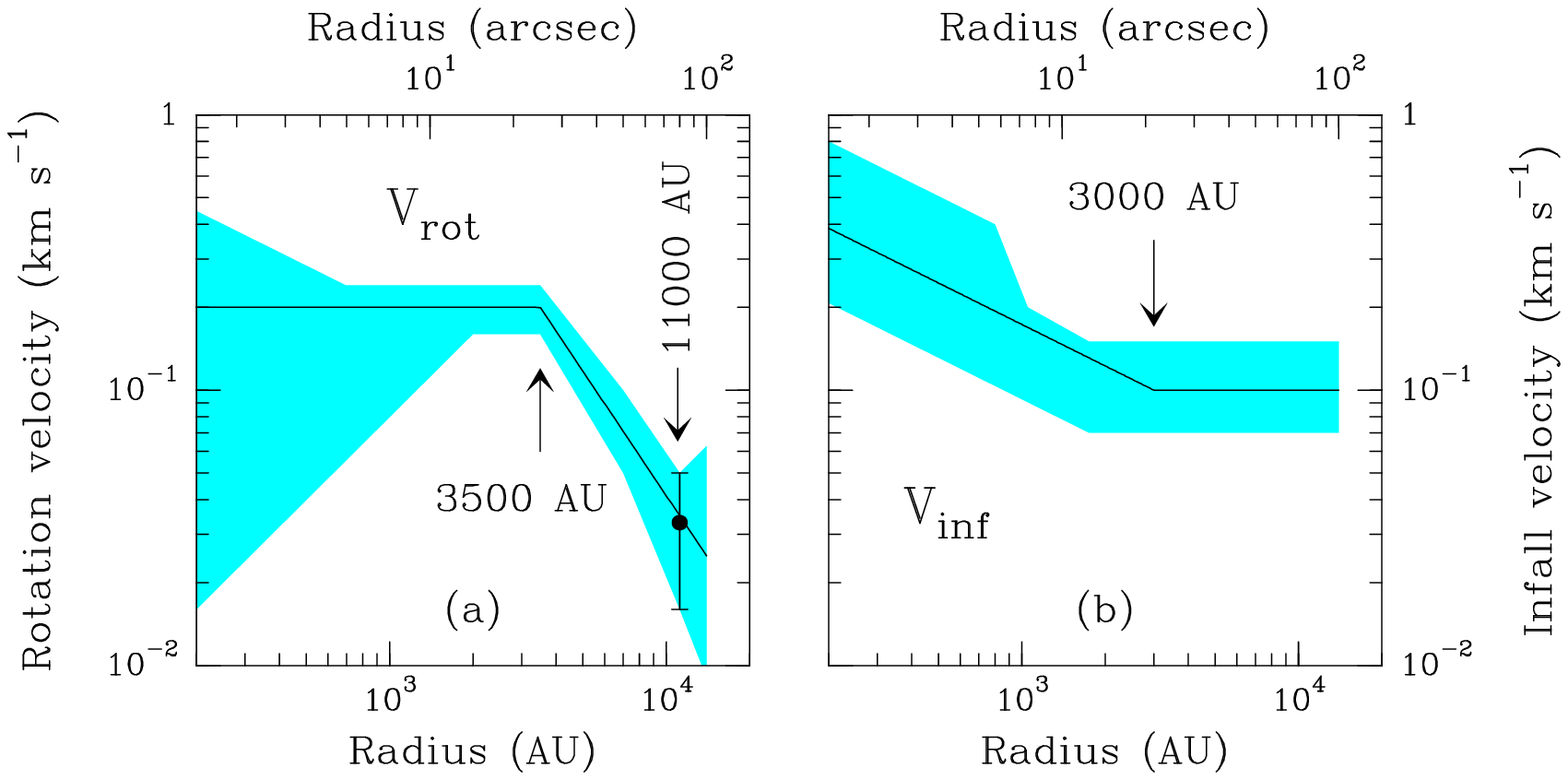,width=65truemm,angle=270,clip=}}}
\captionb{3}{Rotational velocity (a) and infall velocity (b) 
inferred in the IRAM~04191 envelope based on radiative transfer modeling 
of multi-transition CS and C$^{34}$S observations with the
IRAM 30m telescope (Belloche et al. 2002).
The shaded areas show the estimated domains where the 
models match the observations.}
\vskip2mm
\label{iram04191_model}
\end{figure}

Another Class~0 object whose kinematics has been quantified in detail 
is IRAS~4A in the NGC~1333 {\it protocluster} (Di Francesco et al. 2001). 
Using the IRAM Plateau 
de Bure interferometer, Di Francesco et al. observed inverse P-Cygni 
profiles in H$_2$CO($3_{12} - 2_{11}$) toward IRAS~4A, from 
which they derived a very large mass infall rate of 
$\sim 1.1 \times 10^{-4}$~M$_\odot$~yr$^{-1}$ 
at $r \sim 2000$~AU. Even if a warmer initial gas temperature 
($\sim 20$~K) than in IRAM~04191 and some initial level of turbulence are
accounted for (see Di Francesco et al. 2001), this value of
$\dot{M}_{inf}$ corresponds to more than $\sim 15$ times 
the canonical $a_{eff}^3/G$ value (where 
$a_{eff} \simlt 0.3\, \kms$ is the effective sound speed).  
This very high infall rate results both from a very dense 
envelope (a factor $\sim 12$ denser than a SIS at 10~K -- see Motte \& Andr\'e 2001) 
and a large, supersonic infall 
velocity ($\sim 0.68\, \kms $ at $\sim 2000$~AU -- Di Francesco et al. 2001). 
Evidence for fast rotation in the IRAS4A envelope, producing a velocity 
gradient as high as $\sim 40\, \kms \, $pc$^{-1}$, was also reported by 
Di Francesco et al. (2001).

\sectionb{4}{CONCLUSION: COMPARISON WITH COLLAPSE MODELS}

In the case of {\it isolated dense cores} such as those of Taurus, 
the SIS model of Shu~(1977) approximately describes some 
global features of the collapse 
(e.g. the mass infall rate within a factor $\sim 2$) 
but fails to reproduce the observations in detail.
The extended infall velocity profiles observed in 
prestellar cores (Lee et al. 2001) and in the very young Class~0 object
IRAM~04191 (Belloche et al. 2002 -- see \S ~3) are indeed inconsistent 
with a pure inside-out collapse picture.  
The {\it shape} of the density profiles observed in 
prestellar cores are well fitted by purely thermal Bonnor-Ebert sphere models, 
but the {\it absolute} values of the densities are suggestive of some 
additional magnetic support (\S ~2.1). The observed infall velocities are
also marginally consistent with isothermal collapse models starting from 
Bonnor-Ebert spheres (e.g. Foster \& Chevalier 1993, Hennebelle et al. 2003),
as such models tend to produce somewhat faster velocities.
This suggests that the collapse of `isolated' cores is essentially 
{\it spontaneous} and somehow moderated by magnetic effects in (mildly)
magnetized, non-isothermal versions of Bonnor-Ebert cloudlets. 
Indeed, the contrast seen in Fig.~3 between the  
steeply declining rotation velocity profile and the flat infall velocity 
profile of the IRAM~04191 envelope beyond $\sim 3500$~AU 
is very difficult to account for in the context of non-magnetic collapse 
models. 
In the presence of magnetic fields, on the other hand,
the outer envelope can be coupled to, and spun down by, the (large moment 
of inertia of the) ambient cloud (e.g. Basu \& Mouschovias 1995).
 
Based on a qualitative comparison with the ambipolar diffusion 
models of Basu \& Mouschovias, Belloche et al. (2002) propose that 
the rapidly rotating inner envelope of IRAM~04191 corresponds to a 
magnetically (slightly) supercritical core decoupling from an environment still 
supported by magnetic fields and strongly affected by magnetic braking.
A magnetic field $\sim 60\, \mu$G is required at $3500$~AU 
where $\nhd \sim 1-2 \times 10^5\, \rm{cm}^{-3}$, which is comparable to 
the field strengths recently estimated at such densities by 
Crutcher et al. (2003) in three prestellar cores. 
In this picture, the inner $\sim 3500$~AU radius envelope of IRAM~04191 
would correspond to the effective mass reservoir ($\sim 0.5\, M_\odot $) 
from which the central star is being built. 
Moreover, comparison of these results with the 
rotational characteristics of other 
objects in Taurus (Ohashi et al. 1997) indicates that IRAM~04191 behaves 
in a typical manner and is simply observed particularly soon after protostar 
formation. 
The IRAM~04191 example therefore suggests that {\it the masses of stars 
forming in clouds such as Taurus are largely determined by magnetic 
decoupling effects}.

\vskip1mm
\begin{wrapfigure}{i}[0pt]{61mm}
\centerline{\psfig{figure=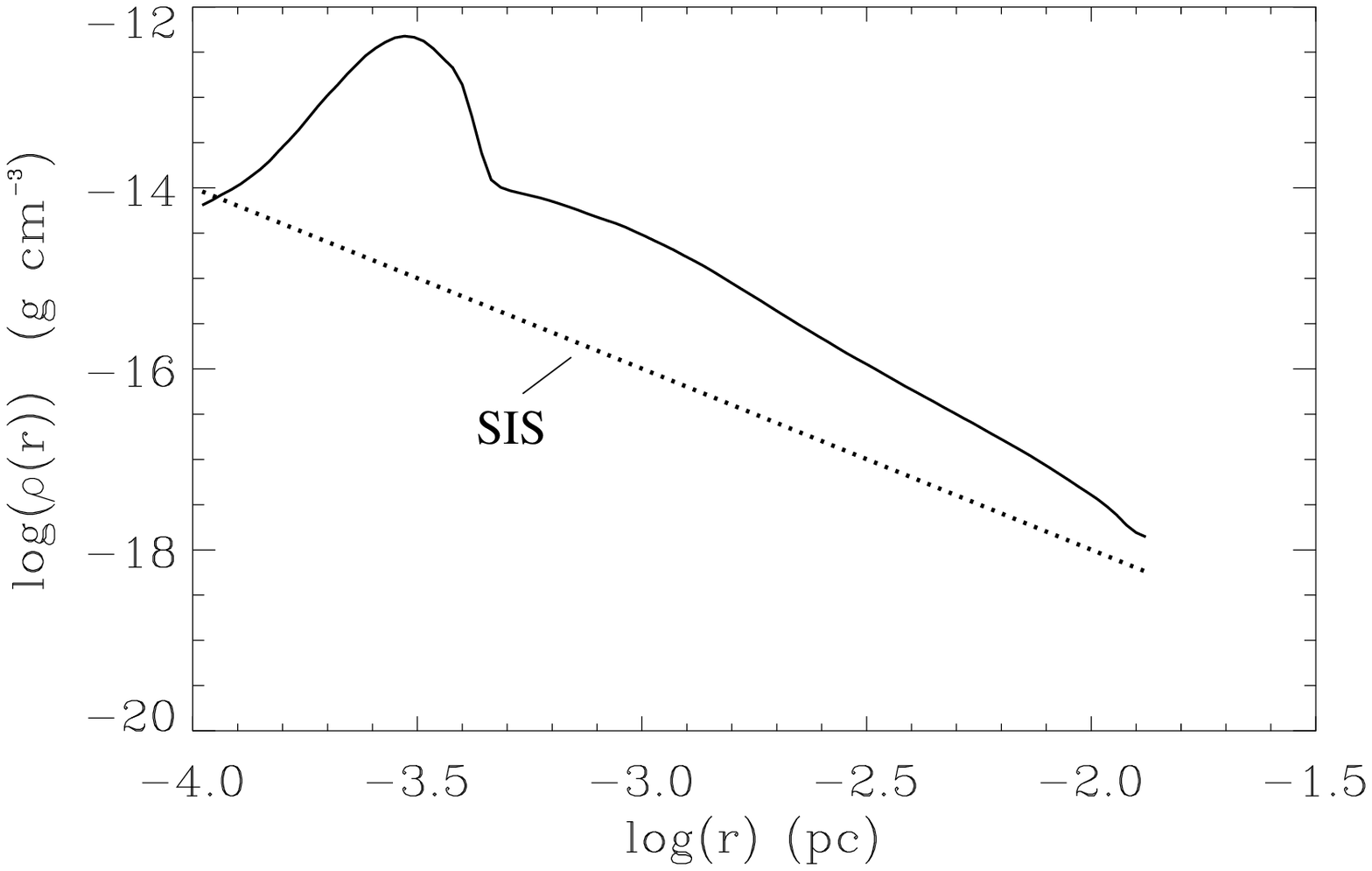,width=65truemm,angle=0,clip=}}
\captionb{4}{Density profile (solid curve) obtained near point mass formation
in SPH simulations of the collapse of an 
initially stable, rotating ($\beta = E_{rot}/E_{grav} = 2\% $) 
Bonnor-Ebert sphere ($T= 10$~K) induced by a very rapid increase
in external pressure (with $P_{ext}/\dot{P}_{ext} = 0.03\ \times $ the initial
sound crossing time) (Hennebelle et al. 2004). 
Note the large overdensity factor
compared to the $\rho \propto r^{-2}$ profile of a SIS at 10~K (dotted line).}
\label{hennebelle}
\end{wrapfigure}

In {\it protoclusters} such as NGC~1333, by contrast, the large overdensity 
factors measured in Class~0 envelopes compared to hydrostatic isothermal 
structures (Motte \& Andr\'e 2001), as well as the fast supersonic
infall velocities and very large infall rates observed in some cases (\S ~3), 
are inconsistent with 
self-initiated forms of collapse and require a {\it strong external influence}.
This point is supported by 
the results of recent
SPH simulations by Hennebelle et al. (2003, 2004). These simulations 
follow the evolution of a Bonnor-Ebert sphere whose collapse
has been induced by an increase in external pressure $P_{ext}$. 
Large overdensity factors (compared to a SIS), together with supersonic 
infall velocities, and large infall rates ($\simgt 10\, \as^3/G$) 
are found near point mass formation 
when (and only when) the increase in $P_{ext}$ 
is strong and very rapid (e.g. Fig.~4), 
resulting in a violent compression wave (cf. Whitworth \& Summers 1985).
Such a violent collapse in protoclusters may be 
conducive to the formation of both massive stars (through higher accretion
rates) and multiple systems (when realistic, non-isotropic compressions are
considered). Future high-resolution studies with the next generation 
of (sub)millimeter instruments (e.g., ALMA -- Wootten 2001) will greatly help test this
view and shed further light on the physics of collapse in cluster-forming regions.

\vskip7mm


\References

\ref
Alves, J.F., Lada, C.J., \& Lada, E.A. 2001, Nature, 409, 159

\ref 
Andr\'e, P., Motte, F., Bacmann, A. 1999, ApJL, 513, L57


\ref 
Andr\'e, P., Ward-Thompson, D., Barsony, M. 2000, in Protostars and Planets IV,
eds. V.~Mannings, A.P.~Boss, \& S.S. Russell (Univ. of Arizona Press, Tucson), p. 59

\ref
Andr\'e, P., Ward-Thompson, D., Motte, F. 1996, A\&A, 314, 625


\ref 
Bacmann, A., Andr\'e, P., Puget, J.-L., Abergel, A., Bontemps, S., 
\& Ward-Thompson, D. 2000, A\&A, 361, 555

\ref
Ballesteros-Paredes, J., Klessen, R.S., \& V\'azquez-Semadeni, E. 2003, ApJ 
592, 188

\ref 
Basu, S., \& Mouschovias, T.Ch. 1995, ApJ, 453, 271 


\ref
Belloche, A., André, P., Despois, D. \& Blinder, S. 2002, A\&A, 393, 927






\ref
Ciolek, G.E., \& Basu, S. 2001, in From Darkness to Light, Eds. T. Montmerle 
\& P. André, ASP Conf. Ser., 243, p. 79

\ref
Ciolek, G.E., Mouschovias, T.Ch. 1994, ApJ, 425, 142

\ref
Crutcher, R.M. 1999, ApJ, 520, 706

\ref
Crutcher, R.M., \& Troland, T.H. 2000, ApJ, 537, L139

\ref
Crutcher, R.M., Nutter, D.J., Ward-Thompson, D., \& Kirk, J. 
et al. 2003, ApJ (astro-ph/0305604)

\ref 
Curry, C.L., \& McKee, C.F. 2000, ApJ, 528, 734

\ref
Di Francesco, J., Myers, P.C., Wilner, D.J. \& Ohashi, N. 2001, ApJ, 562, 770



\ref
Evans, N.J. 1999, ARA\&A, 37, 311

\ref
Evans, N.J. II 2003, in Chemistry as a Diagnostic of Star Formation, 
Eds. C.L. Curry \& M. Fich, in press (astro-ph/0211526)



\ref
Foster, P.N., Chevalier, R.A. 1993, ApJ 416, 303 (FC93)



\ref
Goodman, A.A., Benson, P.J., Fuller, G.A., Myers, P.C. 1993, ApJ, 406, 528




\ref
Hennebelle, P., Whitworth, A.P., Gladwin, P.P., \& André, P. 2003, MNRAS, 340, 870

\ref
Hennebelle, P., Whitworth, A.P., Cha, S., Goodwin, S., 2004, MNRAS, 
in press (astro-ph/0311219)





\ref
Jessop, N., \& Ward-Thompson, D. 2000, MNRAS, 311, 63


\ref Johnstone, D., Wilson, C. D., Moriarty-Schieven, G., et al. 2000, ApJ, 
545, 327

\ref
Jones, C.E., \& Basu, S. 2002, ApJ, 569, 280



\ref
Klessen, R.S., Heitsch, F., \& Mac Low, M.-M. 2000, ApJ, 535, 887





\ref
Lai, S.-P., Velusamy, T., Langer, W.D., \& Kuiper, T.B.H. 
et al. 2003, AJ, 126, 311







\ref
Lee, C.W., Myers, P.C., \& Tafalla, M. 2001, ApJS, 136, 703




\ref
Li, Z.-Y., Shu, F.H. 1997, ApJ, 475, 237














\ref  
Motte, F., Andr\'e, P. 2001, A\&A, 365, 440

\ref  
Motte, F., Andr\'e, P., Neri, R. 1998, A\&A, 336, 150



\ref
Mouschovias, T.C., \& Ciolek, G.E. 1999, in The Origin of Stars and
Planetary Systems, Eds. C.J. Lada \& N.D. Kylafis, Kluwer, p.~305











\ref
Ohashi, N., Hayashi, M., Ho, P.T.P., Momose, M., Tamura, M., Hirano, N., 
Sargent, A. 1997, ApJ, 488, 317







\ref
Padoan, P., \& Nordlund, A. 2002, ApJ, 576, 870





\ref 
Shirley, Y., Evans II, N.J., Rawlings, J.M.C., Gregersen, E.M. 2000, ApJS, 
131, 249

\ref
Shu, F.H. 1977, ApJ, 214, 488

\ref
Shu, F.H., Adams, F.C., Lizano, S. 1987, ARA\&A, 25, 23

\ref
Shu, F.H., Li, Z.-Y., Allen, A. 2003, preprint (astro-ph/0311426)

\ref 
Siebenmorgen, R., \& Kr\"ugel, E. 2000, A\&A, 364, 625





\ref 
Testi, L., Sargent, A.I 1998, ApJL, 508, L91








\ref
Ward-Thompson, D., Andr\'e, P., \& Kirk, J.M. 2002, MNRAS, 329, 257

\ref 
Ward-Thompson, D., Motte, F., Andr\'e, P. 1999, MNRAS, 305, 143

\ref  
Ward-Thompson, D., Scott, P.F., Hills, R.E., Andr\'e, P. 1994, MNRAS, 268, 276




\ref
Whitworth, A., Summers, D. 1985, MNRAS, 214, 1



\ref Wootten, A. 2001, Science with the Atacama Large Millimeter Array, 
ASP Conf. Ser., Vol. 235


\end{document}